\documentclass[11pt,a4paper]{article}
\usepackage[utf8]{inputenc}

\usepackage{epic,eepic,epsfig,amsmath,amssymb,amsthm}
\usepackage{t1enc}
\usepackage{array}

\usepackage[francais,english]{babel}

\usepackage{caption}

\usepackage{color}

\usepackage{bbm}

\usepackage{hyperref}

\def\virgp{\raise 2pt\hbox{,}}

\renewcommand{\geq}{\geqslant}
\renewcommand{\leq}{\leqslant}
\def\N{{\mathbb N}}

\def\R{{\mathbb R}}

\def\virgp{\raise 2pt\hbox{,}}
\def\cdotpv{\raise 2pt\hbox{;}}

\def\1{\mathbbm{1}}

\newtheorem{theorem}{Theorem}[section]
\newtheorem{corollary}[theorem]{Corollary}
\newtheorem{proposition}[theorem]{Proposition}

\theoremstyle{remark}
\newtheorem{remark}{Remark}[section]

\theoremstyle{definition}
\newtheorem{definition}{Definition}[section]

\newtheorem*{notation}{Notation}

\theoremstyle{definition}

\theoremstyle{definition}

\newcolumntype{M}[1]{>{\centering}m{#1}}

\begin{document}

\title{Input-Output Analysis: New Results From Markov Chain Theory (To appear in Economie Appliquée Review)}

\author{Nizar Riane$^\dag$, Claire David$^\ddag$}

\maketitle
\centerline{$^\dag$ Universit\'e Mohammed V de Rabat, Maroc\footnote{Nizar.Riane@gmail.com} }
\vskip 0.5cm

\centerline{$^\ddag$ Sorbonne Universit\'e}

\centerline{CNRS, UMR 7598, Laboratoire Jacques-Louis Lions, 4, place Jussieu 75005, Paris, France\footnote{Claire.David@Sorbonne-Universite.fr}}

\begin{abstract}
In this work, we propose a new lecture of input-output model reconciliation Markov chain and the dominance theory, in the field of interindustrial poles interactions.\\

A deeper lecture of Leontieff table in term of Markov chain is given, exploiting spectral properties and time to absorption to characterize production processes, then the dualities local-global/dominance-Sensitivity analysis are established, allowing a better understanding of economic poles arrangement. An application to the Moroccan economy is given.
\end{abstract}

\maketitle

\vskip 1cm

\noindent \textbf{Keywords}: Graph - Markov chain - Leontieff matrix - Dominance theory.

\vskip 1cm

\noindent \textbf{AMS Classification}: 05C20 - 15A15 - 60J10

\vskip 1cm

\noindent \textbf{JEL Classification}: C65 - C67 - L00

\vskip 1cm

\section{Introduction}

\hskip 0.5cm In his seminal work of~1967, C. Ponsard~\cite{Ponsard1967} proposed to apply graph theory to the analysis of interregional economic flows. Five years later, R. Lantner~\cite{Lantner1972} introduced his economic dominance theory based on Leontieff input-output model, where interdependencies are at stake. Under the Leontieff prism, for instance, a translation (of an interaction) enables one to measure the vulnerability of each patner.\\

  The economic dominance theory makes use of input-output matrices, the coefficients of which correspond to quantitative measures of dependence versus interdependance. One can make a connection between minimal values of the determinant of those matrices, and situations of complete autarky,while maximal values correspond to a situation of perfect dominance. Yet, one thus misses local tools, which would enable one to better understand constantly changing situations.\\

  In 1982, an analogy was made by B. Peterson and M. Olinick in~\cite{Peterson1982} between Leontieff models and Markov chains, by considering the input-output matrix as a transition probability one. Thus, a stationary production vector is related to the case of a closed (no final expenditure) model, while a  productivity condition occurs in the situation of open model. The analysis brought by the authors contain significant algebraic results, but the economic implications are not so clear ; moreover, the link between Markov chains and input-output analysis is not completely exploited.\\

  Recently, D. Lebert~\cite{Lebert2019} generalized this theory to such fields as: international trade in industrial goods, the productive structuring of companies in the United States
and in Western Europe, the economics of innovation and territorial cognitive dynamics.\\

We presently propose to revisit the input-output model as a Markov chains process. We then establish  local results, in term of sensitivity analysis. Our study relies on the same assumptions that can be found in the work of~R.~Lantner~\cite{Lantner1972}: 

\begin{itemize}
\item[\emph{i}.] Homogeneity of poles activity.

\item[\emph{ii}.] No substitution phenomena.

\item[\emph{iii}.]  Constant returns to scale.
\end{itemize}

 Our results establish a close relationship between \emph{dominance phenomena} and \emph{spillover effects}.

\vskip 1cm

\section{Structural analysis : the economic dominance school}

\hskip 0.5cm The theory of dominance, as a branch of structural analysis theory, is based on influence graphs. By representing the interaction web between economic poles, one can thus study the interdependence between poles by extracting, from the resulting graph, the intrinsic underlying information.\\

Starting from the following observation~\cite{Lantner1972}: ``The importance of a transaction between a supplier and a requestor is measured less by its absolute value than by the degree of vulnerability it implies for one or the other'', a global measure of dominance might, first, be deduced from the influence graph.\\

Further developments can be reached then: according to R. Lantner~\cite{Lantner1972}, the macroscopic analysis of the structure as a whole can be achieved by means of the determinant of the input-output matrix. The resulting analysis yields structural indicators, such as: autarky (or independence), hierarchy (or dependence), circularity (or interdependence).\\

In~\cite{Lantner2015}, the author distinguishes three ways to design dominance analysis:

\begin{enumerate}

\item[\emph{i}.]  By respecting the linearity assumption, one can dissociate the direct effect of dominance from indirect ones, and specify the importance of one in relation to the other, as established b~ R.~Lantner and~D.~Lebert in~\cite{LantnerLebert2015}. The stability assumption thus makes it possible to forecast the development of the activity of divisions for all the moments to come. 

\item[\emph{ii}.]  The second approach consists in selecting a certain number of indicators associated to the arrangement of the structure. For instance, it was shown in~\cite{Lantner2001} that the more numerous and more intense were the short circuits in a structure, the weaker was the determinant of its representative matrix. In~\cite{LebertElYounsi2015}, the authors exploit this property to construct and analyze the changes induced on a small number of global indicators by the elimination of some studied poles.

\item[\emph{iii}.]  The third approach exploits properties of Boolean matrices, associated with graphs valuable for studying multiple properties of sets. In~\cite{ElYounsi2015}, H. El Younsi et al. thus analyze the matrix of the world's largest firms holding blocks of technological knowledge, which enable them to enlighten complex multinational strategies associated to the web of patents.

\end{enumerate}

\vskip 1cm

\section{Elements of graph theory and Markov chains}

\hskip 0.5cm An \emph{input-output} table can be understood as a flow matrix, where the flow corresponds to the production transfer from one pole to another. Equivalently, it may also correspond to a demand between poles. With an appropriate normalization, one may transform those flows into transition probabilities between states (poles).\\

  For the sake of clarity, we first recall definitions and results coming from graph theory and Markov chain processes. We refer  to~\cite{Feller1968},~\cite{Harary1994},~\cite{Diestel2017},~\cite{Kemeny1983}, and~\cite{Levin2008} for further details.

\vskip 1cm

\subsection{Graph topology}

\begin{notation}
	
	In the sequel, we denote by~$n$ a strictly positive integer.
	
\end{notation}	
	\vskip 1cm

\begin{definition}[\textbf{Multidigraph~}\cite{Diestel2017}] $\, $\\
		
\hskip 0.5cm A \textbf{multidigraph} $G$ is an ordered pair $G = (V, A)$ where:
\begin{enumerate}
\item $V=\{v_1,\hdots,v_n\}$ is a set of \textbf{vertices},
\item $A\subset V^2$ is a multiset of ordered pairs of vertices, called \textbf{arcs}.
\end{enumerate}

  The graph $G = (V, A, w)$ dotted with a weight function $w \, : \, A \rightarrow \R$ is called a \textbf{weighted multidigraph}.
\end{definition}

\vskip 1cm

\begin{definition}[\textbf{Graph Topology}~\cite{Harary1994}] $\, $\\
	
  Let us denote by~$G =(V, A)$ a multidigraph. We define:

\begin{enumerate}
\item[\emph{i}.] A \textbf{walk} in the multidigraph~$G$ as an alternating sequence of vertices and arcs,~\mbox{$v_0, a_1,v_1,\hdots,a_n,v_n$}, where, for~\mbox{$1 \leq i \leq n$}:
$$a_i=v_{i-1}\, v_i$$

 The length of such a walk is $n$, which is also equal to the number of arcs.\\
\item[\emph{ii}.]  A \textbf{closed walk} as a one with the same first and last points ; a \textbf{spanning walk} as a one that contains all the points.
\item[\emph{iii}.]  A \textbf{path} as a walk where all points are distinct.
\item[\emph{iv}.]  A \textbf{cycle} as a nontrivial closed walk where all points are distinct, except the first and last ones.\\
\end{enumerate}

Given~\mbox{$(i,j)\, \in\,\left \lbrace 1, \hdots, n \right \rbrace^2$}, we will say that:
\begin{enumerate}
	\item[\emph{i}.]  A vertex~$v_j$ is \textbf{accessible} from a vertex $v_i$ if there exists a path connecting $v_i$ to $v_j$. The vertices~$v_j$ and~$v_i$ are then said \emph{mutually accessible}. The \textbf{distance}~\mbox{$d(v_i, v_j)$} from $v_i$ to $v_j$ is equal to the length of any such shortest path.
\item[\emph{ii}.]  If $v_i$ and $v_j$ are mutually accessible, we say they \textbf{communicate}.\\
\end{enumerate}

We will also say that:

\begin{enumerate}
\item[\emph{i}.]  A multidigraph is \textbf{strongly connected} or \textbf{strong} if both vertices of any pair of points
communicate.
\item[\emph{ii}.]  A digraph is \textbf{unilaterally connected} or \textbf{unilateral} if, for any
two points at least one communicate with the other.
\item[\emph{ii}.]  A digraph is \textbf{disconnected} if it is not even unilaterally connected.
\end{enumerate}
\end{definition}

\vskip 1cm

\begin{theorem}[\textbf{\cite{Harary1994}}]$\, $\\

\hskip 0.5cm A digraph is strong if and only if it has a spanning closed walk, and it is unilateral if and only if it has a spanning walk.\\

\end{theorem}

\vskip 1cm

  \emph{Communication between vertices induce an equivalence relation and equivalence classes:}

\vskip 1cm

\begin{definition}[\textbf{Strong component of a graph \cite{Harary1994}}] $\, $\\
		
\hskip 0.5cm Let us consider a multidigraph~$G =(V, A)$. We define:
\begin{enumerate}
\item[\emph{i}.] A \textbf{strong component} of a digraph as a maximal strong subgraph.
\item[\emph{ii}.]  A \textbf{unilateral component} as a maximal unilateral subgraph.
\end{enumerate}

Now, given the strong components~\mbox{$S_1,\hdots, S_n$} of $G$, The \textbf{condensation} $G^{\star}$ of $G$ has the strong components of $G$ as its points, with an arc from $S_i$ to $S_j$ whenever there is at least one arc in $G$ from a point of $S_i$, to a point in $S_j$.
 
\end{definition}

\vskip 1cm

  Graphs are directly related to matrices.

\vskip 1cm
\newpage

\begin{definition}[\textbf{Graph matrices \cite{Harary1994}}] $\, $\\

\hskip 0.5cm Given a multidigraph~\mbox{$G =(V, A)$}, we define:
\begin{enumerate}
\item[\emph{i}.]  The \textbf{adjacency matrix}~\mbox{$\mathbf{A}= \left ( a_{ij}\right)_{1\leq i\leq n,\, 1 \leq j \leq n}$} of $G$ as the square~\mbox{$n \times n$} matrix such that, for any pair of integers~\mbox{$(i,j)\, \in\, \left \lbrace 1,\hdots, n \right \rbrace^2$}:

$$a_{ij} = \left \lbrace \begin{array}{ccc}
	1  & \text{if} & v_iv_j\text{ is an arc of $G$}\\
	0  & \text{otherwise} \end{array}
 \right.$$
 
\item[\emph{ii}.]  The \textbf{accessibility matrix}~\mbox{$\mathbf{R}= \left ( r_{ij}\right)_{1\leq i\leq n,\, 1 \leq j \leq n}$} of $G$ as the square~\mbox{$n \times n$} matrix such that, for any pair of integers~\mbox{$(i,j)\, \in\, \left \lbrace 1,\hdots, n \right \rbrace^2$}:

$$r_{ij} = \left \lbrace \begin{array}{ccc}
	1  & \text{if} &  v_j\text{ is accessible from $v_i$}\\
	0  & \text{otherwise} \end{array}  \right.$$

\item[\emph{iii}.] The \textbf{distance matrix}~\mbox{$\mathbf{D}= \left ( d_{ij}\right)_{1\leq i\leq n,\, 1 \leq j \leq n}$} of $G$ as the square~\mbox{$n \times n$} matrix containing the distances, taken pairwise, between the elements of~$G$. For any pair of integers~\mbox{$(i,j)\, \in\, \left \lbrace 1,\hdots, n \right \rbrace^2$}:

$$d_{ij} = \left \lbrace \begin{array}{ccc}
	d \left ( v_i,v_j\right)  & \text{if} &  v_j\text{ is accessible from $v_i$}\\
	0  & \text{otherwise} \end{array}  \right.$$

\end{enumerate}
\end{definition}

\vskip 1cm

\begin{notation}[\textbf{Hadamard product}] $\, $\\
	
	Given a pair of strictly positive integers~\mbox{$(p,q)$}, and two matrices~\mbox{$M_1= \left ( m_{1,ij}\right)_{1\leq i\leq p,\, 1 \leq j \leq q}$},\\~\mbox{$M_2= \left ( m_{2,ij}\right)_{1\leq i\leq p,\, 1 \leq j \leq q}$}, we denote by~\mbox{$M_1\ast M_2$} their \emph{Hadamard product}, which yields the~\mbox{$p \times q$} matrix:
	
	$$M_1\ast M_2= \left ( m_{1,ij}\, m_{2,ij} \right)_{1\leq i\leq p,\, 1 \leq j \leq q} \, \cdot$$
	
\end{notation}

\vskip 1cm

\begin{theorem}[\textbf{\cite{Harary1994}}] $\, $\\

\hskip 0.5cm Given a multidigraph~\mbox{$G =(V, A)$}, we define, for any pair of integers~\mbox{$(i,j)\, \in\, \left \lbrace 1,\hdots, n \right \rbrace^2$}, the entry~$\mathbf{A}^n_{ij}$ as the number of walks of length $n$ from $v_i$ to $v_j$.


The entries of the accessibility and distance matrices can
be obtained from the powers of $\mathbf{A}$ as follows:
\begin{enumerate}
\item[i.]  For~\mbox{$1 \leq i \leq n$}:
$$\mathbf{R}_{ii} = 1 \quad \text{and} \quad  \mathbf{D}_{ii} = 0$$

\item[ii.]  For~\mbox{$1 \leq i,j \leq n$}:

$$r_{ij} = \left \lbrace \begin{array}{ccc}
1 & \text{if and only if there exist a value of~$n$ such that~$\mathbf{A}^n_{ij} > 0$}\\
	0  & \text{otherwise} \qquad \end{array}  \right.$$

and: 
 $$d_{ij} = \left \lbrace \begin{array}{ccc}
 \displaystyle 	 \min_{k\, \in\, \N^\star} \left  \lbrace \mathbf{A}^k_{ij} > 0 \right \rbrace & & \\
 	+\infty  & \text{otherwise} &\qquad \end{array}  \right.$$
  
\end{enumerate}

 For~\mbox{$1 \leq i \leq n$}, the strong component of $G$ which contains~\mbox{$v_i$} is determined by the entries of~$1$ in the~\mbox{$i^{\text{th}}$} row (or column) of the matrix~\mbox{$\mathbf{S} =\mathbf{R} \ast \mathbf{R}^T$}.
 
\end{theorem}

\vskip 1cm

\subsection{Absorbing Markov chains}

\hskip 0.5cm  In the sequel, we exclusively consider finite Markov chains.

\vskip 1cm

\begin{notation}
	
In the sequel,~$V$ denotes a finite subset of~$\N$.

\end{notation}
	
	\vskip 1cm

\begin{definition}[\textbf{Finite Markov chain \cite{Bremaud2008}}] $\, $\\

\hskip 0.5cm A sequence $(X_n)_{n\geq 0}$ of random variables is a \textbf{finite Markov chain} with finite state space $V$ and transition matrix $\mathbf{P}$ if, for all states $i_0,\hdots,i_{n-1},i,j \in V$, and any integer~$n \geq 1$, one has:

\begin{align*}
P(X_{n+1}=j \, | \, X_n=i,X_{n-1}=i_{n-1},\hdots,X_0=i_0)&=P(X_{n+1}=j \, | \, X_n=i)=\mathbf{P}_{ij}
\end{align*}

  The transition matrix $\mathbf{P}$ is \textbf{stochastic}, in the sense that its entries are all non-negative and such that:

$$\forall \, i\, \in\, V\, : \quad \displaystyle
\sum_{j\in V} \mathbf{P}_{ij} =1
$$

\end{definition}

\vskip 1cm

\begin{definition}[\textbf{Stochastic and substochastic matrices}]$\, $\\

\hskip 0.5cm A~$n\times n$ matrix with non negative entries is said to be \textbf{stochastic} (resp. \textbf{substochastic}) if, for any pair of integers~\mbox{$(i,j)\, \in\, \left \lbrace 1,\hdots, n \right \rbrace^2$}, one has:

$$ \displaystyle 
\sum_{j=1}^n \mathbf{P}_{ij}=1 \quad \left (\text{resp. } \sum_{j=1}^n \mathbf{P}_{ij}\leq 1 \right)$$
\end{definition}

\vskip 1cm

One may now establish an analogy between graphs and Markov chains.

\vskip 1cm

\begin{definition}[\textbf{Random walk on a graph}] $\, $\\

\hskip 0.5cm Given a weighted multidigraph~\mbox{$G = (V,A,w)$}, and a stochastic~\mbox{$n \times n$} matrix~$\mathbf{P}$, we define a \textbf{random walk} on $G$ as the Markov chain with transition matrix 
$$\mathbf{P}=\left ( w_{ij} \right)_{1\leq i,j\leq n}$$

Conversely, given a Markov chain on the finite state space $V$, one may build a weighted multidigraph~\mbox{$G = (V,A,w)$}, whose vertices set is the state space $V$, while the  weighted arcs are defined by the transition probabilities:

$$\forall\, (i,j)\, \in\, \left \lbrace 1,\hdots, n \right \rbrace^2\, : \quad P(X_{k+1}=j \, | \, X_k=i) \quad , \quad  k\geq 0$$

\end{definition}

\vskip 1cm

 In a similar way, one may define accessibility and communication between states:

\vskip 1cm

\begin{definition}[\textbf{Accessibility and communication \cite{Feller1968} }]$\, $\\

\hskip 0.5cm Let us consider a random walk~\mbox{$(X_n)_{n\geq 0}$} on $G=(V,A)$, with transition matrix $\mathbf{P}$. Then:
\begin{enumerate}
\item[\emph{i}.] A state~\mbox{$j\,\in\, V$} is said  \textbf{accessible} from another one~\mbox{$i\,\in\, V$} if the exists a natural integer~$k$ such that~\mbox{$\mathbf{P}^k_{ij}>0$}. This is equivalent to the fact that $G$ contains a directed path from $v_i$ to $v_j$. \\

Such a relation will de denoted as:

$$ i \rightarrow j$$

\item[\emph{ii}.]  A state~\mbox{$j\,\in\, V$} \textbf{communicates} with another one~\mbox{$i\,\in\, V$} if:
$$ i\rightarrow j \quad \text{and} \quad j\rightarrow i$$

	Such a relation will de denoted as:
	
$$ i \leftrightarrow j$$

\item[\emph{iii}.] The relation $\leftrightarrow$ is a equivalence one.
\item[\emph{iv}.]  A subset $C\subset V$ is \textbf{closed} if no state outside $C$ is accessible from any state in $C$.
\item[\emph{v}.] A Markov chain is \textbf{irreducible} if it contains a unique closed set.
\item[\emph{vi}.]  If, given a state~\mbox{$i\,\in\, V$}  such that~\mbox{$\mathbf{P}_{ii}=1$}, the state $i$ is said to be \textbf{absorbing}.
\item[\emph{vii}.] Given a state~\mbox{$i\,\in\, V$}, the $gcd \{k \,\in\,\N^\star\, :\quad  \mathbf{P}^k_{ii} > 0\}$, where $gcd$ denotes the greatest common divisor, is the \textbf{period} of the state $i$. If $d = 1$, the state $i$ is \textbf{aperiodic}.

\end{enumerate}

\end{definition}

\vskip 1cm

\begin{remark}

\noindent Closed sets and strong components are two different notions: a closed set could be disconnected and a strong component could be open.\\

One may also note that:

$$\# \{ i \, \in \, V \, : \, \mathbf{P}^k_{ii}\neq 0\}$$

\noindent  represent the number of closed walks of length~$k$.

\end{remark}

\vskip 1cm

\begin{definition}[\textbf{Irreducible Matrix}] $\, $\\
	
A matrix~$A$ is \textbf{irreducible} if it is not similar via a permutation to a block upper triangular matrix. In the case of the  adjacency matrix of a directed graph, it is irreducible if and only if the graph is strongly connected.
	
\end{definition}

\vskip 1cm

\begin{theorem}[\textbf{\cite{Feller1968}} ]$\, $\\
		
\hskip 0.5cm	Given a Markov transition~$\mathbf{P}$ matrix on a finite state~$V$, the restriction of $\mathbf{P}$ to a closed set~\mbox{$C\subset V$} is also a Markov chain.
\end{theorem}

\vskip 1cm

\begin{definition}[\textbf{States classification \cite{Feller1968}}] $\, $\\
		
\hskip 0.5cm  For~\mbox{$1 \leq i ,j\leq n$}, let us denote by~\mbox{$f_{ij}^n$} the probability that, starting from the state~$i$, the process will pass through~$j$ at the~\mbox{$n^\text{th}$ step}. We introduce:

\begin{enumerate}
	\item[\emph{i}.]  The probability that starting from the state $i$ the system will ever pass through $j$:
	
	$$f_{ij} =\displaystyle \sum_{n=1}^{+\infty}f_{ij}^n $$

	\item[\emph{ii}.]  The mean recurrence time for $i$:
	
	$$ \displaystyle    \nu_i=\sum_{n=1}^{+\infty}nf_{ij}^n
	$$
	
\end{enumerate}

 We will say that:

\begin{enumerate}
\item[\emph{i}.] The state~$i$ is \textbf{persistent} if~\mbox{$f_{ii}=1$}.
\item[\emph{ii}.] The state~$i$ is \textbf{transient} if~\mbox{$f_{ii}<1$}.
\end{enumerate}

\end{definition}

\vskip 1cm

\begin{theorem}[\textbf{\cite{Feller1968}} ] $\,$\\

\hskip 0.5cm The states of a finite Markov chain can be divided, in a unique manner, into non-overlapping sets $T,C_1,\hdots, C_q$ such that:

\begin{enumerate}
	\item[i.] $T$ consists of all transient states.
	\item[ii.]  There exists at least one closed set~\, $C_k\subset V$.
	\item[iii.]  If $i\, \in\,  C_k$, then:
	
	 $$\forall\, j\, \in\, C_k\, : \quad f_{ij}=1 \quad \text{while} \quad 
	 \forall\, j\, \notin\, C_k\, : \quad f_{ij}=0$$
\end{enumerate}
\end{theorem}

\vskip 1cm

\newpage
\begin{definition}[\textbf{Absorbing Markov chain \cite{Kemeny1983} }]$\,$\\

\hskip 0.5cm A Markov chain is said to be \textbf{absorbing} if all of its non-transient states are absorbing.\\

 The transition matrix~$\mathbf{P}\in M_n(\R)$ of a finite absorbing Markov chain can be represented in a canonical form:

\begin{align*}
\mathbf{P}&=\left(
\begin{matrix}
\mathbf{B} & \mathbf{R} \\
\mathbf{0} & \mathbf{I} \\
\end{matrix}
\right)
\end{align*}

\noindent where $\mathbf{B}\in M_p(\R)$, $\mathbf{B}\in M_{pq}(\R)$ and $\mathbf{I}\in M_q(\R)$, and $p$ (resp. $q$) is the cardinal of transient (resp. absorbant) states.

\end{definition}

\vskip 1cm

\begin{proposition}[\textbf{\cite{Kemeny1983}} ]$\, $\\

\hskip 0.5cm Let us consider the stochastic matrix~$\mathbf{P}$ of an absorbing Markov chain. We define the \textbf{fundamental matrix} as:

\begin{align*}
\mathbf{N}&=(\mathbf{I}-\mathbf{B})^{-1}=\sum_{i=0}^{+\infty} \mathbf{B}^i
\end{align*}

Then, for any pair of integers~\mbox{$(i,j)\, \in\, \left \lbrace 1,\hdots, n \right \rbrace^2$}:
\begin{enumerate}
\item[i.] The probability~\mbox{$\mathbf{B}_{ij}$} of being absorbed in the absorbing state $j$ when starting from transient state $i$, is also the coefficient of indices~\mbox{$i,j$} of the product matrix~\mbox{$ \mathbf{N}\,\mathbf{R}$}:

\begin{align*}
\mathbf{B}_{ij}&=\left ( \mathbf{N}\,\mathbf{R}\right)_{ij}
\end{align*}

\item[ii.]  The quantity~\mbox{$\mathbf{N}_{ij}$} represents the expected number of visits to a transient state $j$ starting from a transient state $i$.
\item[iii.] The expected number of steps (time)~\mbox{ $\mathbf{t}_i$} before being absorbed when starting in state $i$, is such that:

\begin{align*}
\mathbf{t}&=\left(
\begin{matrix}
\mathbf{t}_1\\
\vdots\\
\mathbf{t}_{p}
\end{matrix}
\right)=\mathbf{N}\,\left(
\begin{matrix}
 1\\
	\vdots\\
1
\end{matrix}
\right)
\end{align*}


\end{enumerate}

\end{proposition}

\vskip 1cm

\begin{definition}[\textbf{Stationary distribution \cite{Feller1968}}] $\, $\\

\hskip 0.5cm Given a transition matrix $\mathbf{P}$ and a probability distribution~$\mu$ on $V$, we say that~$\mu$ on $V$ is \textbf{stationary} if:
\begin{align*}
\mu \mathbf{P}&=\mu \, \cdot
\end{align*}

\end{definition}

\vskip 1cm

\begin{theorem}[\textbf{Existence and uniqueness of stationary distribution \cite{Feller1968}}] $\, $\\
		
\hskip 0.5cm Every finite Markov chain admit a stationary distribution $\mu$ and the stationary distribution is unique if and only if the chain is irreducible.\\

 Moreover, $\mu(i)=0$ for any transient state~\mbox{$i\, \in\,  V$}.

\end{theorem}

\vskip 1cm

\begin{theorem}[\textbf{Spectrum of a finite Markov chain \cite{Levin2008} }] $\, $\\

\hskip 0.5cm Let us denote by~$\mathbf{P}$ be the transition matrix of a finite Markov chain. Then:

\begin{enumerate}
\item[i.]  If $\lambda$ is an eigenvalue of $\mathbf{P}$, then $|\lambda|\leq1$.
\item[ii.] The eigenvalues of $\mathbf{P}$ of modulus equal to~$1$ are complex roots of unity. The~\mbox{$d^\text{th}$} roots of unity are eigenvalues of $\mathbf{P}$ if and only if $\mathbf{P}$ has a recurrent class with period $d$. The multiplicity of each~\mbox{$d^\text{th}$} root of unity is equal to the number of recurrent classes of period~$d$.
\item[iii.]  If $\mathbf{P}$ is irreducible, the vector space of eigenfunctions corresponding to the eigenvalue $1$ is the one-dimensional space generated by the column vector 
$$\mathbf{1}=(1, \hdots , 1)^T$$
\item[iv.]  If $\mathbf{P}$ is irreducible and aperiodic, then $-1$ is not an eigenvalue of $\mathbf{P}$.
\end{enumerate}
\end{theorem}

\vskip 1cm

\begin{definition}[\textbf{Spectral gap and relaxation time \cite{Levin2008}}] $\, $\\
		
  Let us consider the eigenvalues of transition matrix~$\mathbf{P}$:

  $$-1 \leq \lambda_{n} \leq \hdots \leq \lambda_2 < \lambda_1 = 1  $$

  We set:
   
$$
\lambda^{\star}=\max \{ |\lambda| \, : \, \lambda \text{ is an eigenvalue of } \mathbf{P}, \, \lambda\neq 1\}
$$

Then:
\begin{enumerate}
\item[\emph{i}.] The \textbf{spectral gap} is defined by:~\mbox{$\gamma = 1 -\lambda_2$}.
\item[\emph{ii}.]  The \textbf{absolute spectral gap} is equal to the difference: 
$$\gamma^{\star} = 1 - \lambda^{\star}$$

 If the matrix~$\mathbf{P}$ is aperiodic and irreducible, one has then: ~\mbox{$\gamma^{\star} > 0$}.
\item[\emph{iii}.]  The \textbf{relaxation time} $t_{\text{rel}}$ is defined as:

\begin{align*}
t_{\text{rel}}&=\frac{1}{\gamma^{\star}}
\end{align*} 
\end{enumerate}

\end{definition}

\vskip 1cm

\newpage

Let us now recall an important result about non-negative matrices:

\vskip 1cm

\begin{theorem}[\textbf{General Perron-Frobenius \cite{Horn2013}}]
		$\, $\\ \label{Perron-Frobenius}
	
 Let us consider a non-negative matrix~$\mathbf{A}$, with spectral radius~\mbox{$\rho(\mathbf{A})$}. Then:  

\begin{enumerate}
\item[i.]  $\rho(\mathbf{A})$ is an eigenvalue of~$\mathbf{A}$, and there
exists a non-negative, non-zero vector~$\mathbf{v}$ such that: $$\mathbf{A}\mathbf{v}=\rho(\mathbf{A})\mathbf{v}$$

\item[ii.]~\mbox{$ \displaystyle 
\min_{1\leq i\leq n} \sum_{j=1}^n \mathbf{A}_{ij} \leq \rho(\mathbf{A}) \leq \max_{1\leq i\leq n} \sum_{j=1}^n \mathbf{A}_{ij}$}.

\end{enumerate}

\end{theorem}

\vskip 1cm

\section{Input-output model : a Markov chain formulation}

 \hskip 0.5cm We now consider an economy with~\mbox{$n\, \in\, \N^\star$} poles, where, for~\mbox{$1 \leq i \leq n$}, the production of the~\mbox{$i^{th}$} pole is denoted by~\mbox{$X_i$}, while its final consumption is~\mbox{$Y_i$}.  The production is shared, with intermediary consumption~\mbox{$x_{ij}$} between the~\mbox{$i^{th}$} and~\mbox{$j^{th}$} poles,~\mbox{$1 \leq i,j \leq n$}. The input-output system can be then written as:

$$
\forall \, (i,j)\, \in\, \left \lbrace 1,\hdots,n \right \rbrace^2\, :\quad X_i- \sum_{j=1}^n x_{ij}  = Y_i\\
$$

Such a design is called \textbf{direct orientation} of the flow. The \textbf{indirect} one is obtained by considering the origin (supply) instead of the destination (demand):

 $$
\forall \, (i,j)\, \in\, \left \lbrace 1,\hdots,n \right \rbrace^2\, :\quad X_i- \sum_{j=1}^n x_{ji}  = W_i
$$

\noindent where $W_i$ denotes the added value of the~\mbox{$i^{th}$} pole.\\

  For~\mbox{$1 \leq i,j \leq n$}, we respectively define the \textbf{technical} and \textbf{trade} coefficients, as:

  $$\theta_{ij}=\displaystyle\frac{x_{ij}}{X_j} \quad , \quad \alpha_{ij}=\displaystyle\frac{x_{ij}}{X_i}$$
  
 One may then rewrite the model under the following matrix form:

$$
\mathbf{X}-\mathbf{\Theta}\mathbf{X} =\mathbf{Y} \quad , \quad 
\mathbf{X}^T-\mathbf{X}^T\mathbf{A} =\mathbf{W}^T\\
$$

\noindent where

\begin{align*}
\mathbf{\Theta}=\left(
\begin{matrix}
\theta_{11} & \hdots & \theta_{1n}\\
\vdots & \ddots & \vdots \\
\theta_{n1} & \hdots & \theta_{nn}\\
\end{matrix}
\right), \quad
\mathbf{A}=\left(
\begin{matrix}
\alpha_{11} & \hdots & \alpha_{1n}\\
\vdots & \ddots & \vdots \\
\alpha_{n1} & \hdots & \alpha_{nn}\\
\end{matrix}
\right), \quad
\mathbf{X}=\left(
\begin{matrix}
X_1\\
\vdots\\
X_n
\end{matrix}
\right), \quad
\mathbf{Y}=\left(
\begin{matrix}
Y_1\\
\vdots\\
Y_n
\end{matrix}
\right), \quad
\mathbf{W}=\left(
\begin{matrix}
W_1\\
\vdots\\
W_n
\end{matrix}
\right) \\
\end{align*}

Let us set:

 $$Y=\displaystyle \sum_{i=1}^n Y_i \quad   \text{and} \quad W=\sum_{i=1}^n W_i  $$
 
$Y$ represents the total final expenditure, while~$W$ stands for  the total added value. \\

Let us then introduce the following augmented matrices:

\begin{align*}
\widehat{\mathbf{\Theta}}&=\left(
\begin{matrix}
\mathbf{\Theta} & \mathbf{0}\\
\mathbf{w}^T & 1\\
\end{matrix}
\right),&
\mathbf{w}&=\left(
\begin{matrix}
{W_1}/{X_1}\\
\vdots\\
{W_n}/{X_n}
\end{matrix}
\right),&
\widehat{\mathbf{X}}&=\left(
\begin{matrix}
\mathbf{X}\\
W
\end{matrix}
\right),&
\widehat{\mathbf{Y}}&=\left(
\begin{matrix}
\mathbf{Y}\\
-W
\end{matrix}
\right),
\\
\tilde{\mathbf{A}}&=\left(
\begin{matrix}
\mathbf{A} & \mathbf{y}\\
\mathbf{0} & 1\\
\end{matrix}
\right),&
\mathbf{y}&=\left(
\begin{matrix}
{Y_1}/{X_1}\\
\vdots\\
{Y_n}/{X_n}
\end{matrix}
\right),&
\tilde{\mathbf{X}}&=\left(
\begin{matrix}
\mathbf{X}\\
Y
\end{matrix}
\right),&
\tilde{\mathbf{W}}&=\left(
\begin{matrix}
\mathbf{W}\\
-Y
\end{matrix}
\right)\\
\end{align*}

This yields:

\begin{align*}
\widehat{\mathbf{X}}-\tilde{\mathbf{\Theta}}\widehat{\mathbf{X}}&=\left(
\begin{matrix}
\mathbf{X} \\
W
\end{matrix}
\right) -  
\left(
\begin{matrix}
\mathbf{\Theta} & \mathbf{0}\\
\mathbf{w}^T & 1\\
\end{matrix}
\right) \left(
\begin{matrix}
\mathbf{X} \\
W
\end{matrix}
\right)\\
&=\left(
\begin{matrix}
\mathbf{X} \\
W
\end{matrix}
\right) -  \left(
\begin{matrix}
\mathbf{\Theta}\mathbf{X} \\
\mathbf{w}^T \mathbf{X} + W
\end{matrix}
\right)\\
&=\left(
\begin{matrix}
\mathbf{X} \\
W
\end{matrix}
\right) -  \left(
\begin{matrix}
\mathbf{\Theta}\mathbf{X} \\
2W
\end{matrix}
\right)\\
&=\widehat{\mathbf{Y}}  
\end{align*}

\noindent and:

\begin{align*}
	\tilde{\mathbf{X}}^T-\tilde{\mathbf{X}}^T\tilde{\mathbf{A}}&=\left(
	\begin{matrix}
		\mathbf{X}^T &Y
	\end{matrix}
	\right) - \left(
	\begin{matrix}
		\mathbf{X}^T &Y
	\end{matrix}
	\right) 
	\left(
	\begin{matrix}
		\mathbf{A} & \mathbf{y}\\
		\mathbf{0} & 1\\
	\end{matrix}
	\right)\\
	&=\left(
	\begin{matrix}
		\mathbf{X}^T &Y
	\end{matrix}
	\right) - \left(
	\begin{matrix}
		\mathbf{X}^T \mathbf{A} & \mathbf{X}^T\mathbf{y} + Y
	\end{matrix}
	\right)\\
	&=\left(
	\begin{matrix}
		\mathbf{X}^T &Y
	\end{matrix}
	\right) - \left(
	\begin{matrix}
		\mathbf{X}^T \mathbf{A} & 2 Y
	\end{matrix}
	\right)\\
	&=\tilde{\mathbf{W}}^T \\
\end{align*}

 This artifact enables one to transform the model into an \textbf{absorbing Markov chain} with~\mbox{$n+1$} states, where~\mbox{$\widehat{\mathbf{\Theta}}$} and~\mbox{$\tilde{\mathbf{A}}$} play the role of the transition matrices (one may note that~\mbox{$\widehat{\mathbf{\Theta}}^T$} and~\mbox{$\tilde{\mathbf{A}}$} are stochastic matrices,while~\mbox{$\mathbf{\Theta}^T$} and~\mbox{$\mathbf{A}$} are substochastic ones).

\vskip 1cm

\begin{remark}{\ \label{Eigenvalue} }\\

\hskip 0.5cm It is possible to obtain the final expenditure related to each pole, but with no significant interest in our situation. In the case of the indirect scheme, the matrices have the form:

\begin{align*}
\tilde{\mathbf{A}}&=\left(
\begin{matrix}
\mathbf{A} & \mathbb{Y}\\
\mathbf{0} & \mathbf{I}\\
\end{matrix}
\right),&
\mathbb{Y}&=\left(
\begin{matrix}
{Y_1}/{X_1}&  & 0\\
& \ddots &\\
0 & & {Y_n}/{X_n}
\end{matrix}
\right),&
\tilde{\mathbf{X}}&=\left(
\begin{matrix}
\mathbf{X}\\
\mathbf{Y}
\end{matrix}
\right),&
\tilde{\mathbf{W}}&=\left(
\begin{matrix}
\mathbf{W}\\
-\mathbf{Y}
\end{matrix}
\right)
\end{align*}

\end{remark}

\vskip 1cm

\newpage
  By construction, the substochastic matrices~\mbox{$\mathbf{\Theta}^T$} and~\mbox{$\mathbf{A}$} are such that:

$$\displaystyle 
0 \leq \sum_{i=1}^n \mathbf{\Theta}_{ij} = 1-\mathbf{w}_j \leq 1\quad \text{and} \quad 
0 \leq \sum_{j=1}^n \mathbf{A}_{ij} = 1-\mathbf{y}_i \leq 1
$$

  It follows from Theorem~\ref{Perron-Frobenius} that:

$$\displaystyle 
\min_{1 \leq j \leq n} \left ( 1-\mathbf{w}_j \right)  \leq \rho(\mathbf{\Theta}) \leq \max_{1 \leq j \leq n}  \left ( 1-\mathbf{w}_j \right )\quad \text{and} \quad 
\min_{1 \leq i \leq n}  \left ( 1-\mathbf{y}_i \right)  \leq \rho(\mathbf{A}) \leq \max_{1 \leq i\leq n} \left ( 1-\mathbf{y}_i \right)
$$

\vskip 1cm

  We have thus proved the following result:

\vskip 1cm

\begin{corollary}{\ }\\
	
	One has:
$$\displaystyle 
\lim_{k\rightarrow+\infty} \mathbf{A}^k=\lim_{k\rightarrow+\infty} \mathbf{\Theta}^k=\mathbf{0}$$
	
	\noindent and:
	
	$$
\mathbf{O} =\left(\mathbf{I}-\mathbf{\Theta}\right)^{-1}=\sum_{k=0}^{+\infty}\mathbf{\Theta}^k \geq 0
\quad , \quad 
\mathbf{N} =\left(\mathbf{I}-\mathbf{A}\right)^{-1}=\sum_{k=0}^{+\infty}\mathbf{A}^k \geq 0
$$

\end{corollary}

\vskip 1cm

\begin{remark}{\ }\\
	
\hskip 0.5cm The spectrum of the block triangular matrix~\mbox{$\tilde{\mathbf{A}}$} (resp.~\mbox{$\widehat{\mathbf{\Theta}}$}) is the same as the one of~$\mathbf{A}$ (resp.~$\mathbf{\Theta}$) and~$1$.
\end{remark}

\vskip 1cm

\begin{remark}{\ }\\

Let us consider \emph{the intermediary expenditures matrix}~\mbox{$\mathbf{\Omega}$}, and the diagonal one~$\mathbb{X}$ such that:

$$\forall\, i\, \in\, \left \lbrace 1, \hdots, n \right \rbrace\, : \quad \mathbb{X}_{ii}=X_i$$
 
 One has then:

$$
\mathbf{\Theta} =\mathbf{\Omega}\, \mathbb{X}^{-1} \quad \text{and} \quad 
\mathbf{A} =\mathbb{X}^{-1}\, \mathbf{\Omega}
$$

One might thus deduce that the direct and the indirect schemes have the same connection and spectral properties. Indeed, given an eigenvalue-eigenvector couple~\mbox{$(\lambda,v)$}  of $\mathbf{\Theta}$:

\begin{align*}
\mathbf{\Theta}v&=\mathbf{\Omega}\mathbb{X}^{-1}v=\lambda v
\end{align*}

Multiplication of each side by~$\mathbb{X}^{-1}$ yields:

\begin{align*}
\mathbb{X}^{-1}\mathbf{\Theta}v&=\left(\mathbb{X}^{-1}\mathbf{\Omega}\right)\left(\mathbb{X}^{-1}v\right)=\mathbf{A}\left(\mathbb{X}^{-1}v\right)=\lambda \left(\mathbb{X}^{-1} v\right)
\end{align*}

\noindent which means that~\mbox{$(\lambda,\mathbb{X}^{-1} \, v)$} is the corresponding eigenvalue-eigenvector couple of~$\mathbf{A}$.

\end{remark}

\vskip 1cm

\subsection{Dual interpretation}

\hskip 0.5cm Let us consider again our economy with~\mbox{$n\, \in\, \N^\star$} poles: for~\mbox{$1 \leq i \leq n$}, the production flow~\mbox{$X_i$} of the~\mbox{$i^{th}$} pole is equivalent to a monetary flow in the opposite direction (demand)~\mbox{$M^d_i$}. For~\mbox{$1 \leq i,j \leq n$}, the intermediary expenditure~\mbox{$x_{ij}$} induces an intermediate monetary supply~\mbox{$m_{ji}$}, while the final consumption~\mbox{$Y_i$} induces a final monetary supply~\mbox{$K_i$}. The dual monetary input-output system can thus be written as:

$$
\forall \, (i,j)\, \in\, \left \lbrace 1,\hdots,n \right \rbrace^2\, :\quad 
M^d_i - \displaystyle \sum_{j=1}^n m_{ji}  = K_i\\
$$

 For~\mbox{$1 \leq i,j \leq n$}, we define the \textbf{monetary supply coefficients}:

 $$\gamma_{ij}=\displaystyle\frac{m_{ij}}{M^d_i}$$

One gets then the \textbf{dual monetary problem} in term of monetary flows:

\begin{align*}
\mathbf{M^d}^T(\mathbf{I}-\mathbf{G})&=\mathbf{K}^T\\
\end{align*}

\noindent where

$$
\forall \, (i,j)\, \in\, \left \lbrace 1,\hdots,n \right \rbrace^2\, :\quad \gamma_{ij}=\displaystyle\frac{m_{ij}}{M^d_i}=\displaystyle\frac{x_{ji}}{X_i}=\theta_{ji}$$

\noindent  which means that~\mbox{$\mathbf{G}=\mathbf{\Theta}^T$}. One may then write:

\begin{align*}
\mathbf{M^d}^T(\mathbf{I}-\mathbf{\Theta}^T)&=\mathbf{K}^T\\
\end{align*}

 In particular, the matrix $\mathbf{G}$ is substochastic.

\vskip 1cm

\section{Sensitivity analysis}

\hskip 0.5cm  Using the fundamental matrices~$\mathbf{O}$ and~$\mathbf{N}$, one has:

$$
\mathbf{X}  =\mathbf{O}\mathbf{Y} \quad , \quad 
\mathbf{X}^T =\mathbf{W}^T \mathbf{N}
$$

\noindent where:

$$
\forall \, (i,j)\, \in\, \left \lbrace 1,\hdots,n \right \rbrace^2\, :\quad
\mathbf{X}_i =\displaystyle   \sum_{j=1}^n\mathbf{O}_{ij}\mathbf{Y}_j \quad , \quad 
\mathbf{X}_j =\displaystyle \sum_{i=1}^n \mathbf{W}_i \mathbf{N}_{ij}
$$

\noindent which yield:

$$
\frac{\partial \mathbf{X}_i}{\partial \mathbf{Y}_j} =\mathbf{O}_{ij} \quad , \quad 
\frac{\partial \mathbf{X}_j}{\partial \mathbf{W}_i} = \mathbf{N}_{ij}
$$

  In the monetary sphere, one may write:

$$
\mathbf{M^d}^T =\mathbf{K}^T(\mathbf{I}-\mathbf{G})^{-1}\\
=\mathbf{Y}^T\mathbf{Q}
$$

\newpage
 For~\mbox{$j\, \in\, \{1,\hdots,n\}$}, one has:

\begin{align*}
\mathbf{M^d}_j&=\sum_{i=1}^n \mathbf{Y}_i \mathbf{Q}_{ij}
\end{align*}

 In particular:

\begin{align*}
\frac{\partial \mathbf{M^d}_j}{\partial \mathbf{Y}_i}&= \mathbf{Q}_{ij}\\
\end{align*}

  From the Markov chains theory, one knows that $\mathbf{O}^T_{ij}$, $\mathbf{N}_{ij}$ and $\mathbf{Q}_{ij}$ represent the expected number of visits to~$j$ starting from~$i$. In the specific case of the indirect scheme, may one compute the expected time~\mbox{ $\mathbf{t}_i$ } related to the production of the~\mbox{$i^{th}$} pole before been absorbed by final demand, one gets:

\begin{align*}
\mathbf{t}&=\left(
\begin{matrix}
\mathbf{t}_1\\
\vdots\\
\mathbf{t}_{n}
\end{matrix}
\right)=\mathbf{N}\,\mathbf{1}
\end{align*}

\noindent which is not more than production process duration vector of each pole.

\vskip 1cm

\subsection{Long run effects and relaxation time}

\hskip 0.5cm In the first section, we have introduced the relaxation time as the inverse of the spectral gap, which is the difference between the absolute values of the smallest and the greatest eigenvalue$\lambda^{\star}$.\\

 May one consider the indirect scheme, it follows from remark~\ref{Eigenvalue} that the maximal spectral radius of $\mathbf{A}$ is:

 $$\lambda^{\star}=\displaystyle \max_{1 \leq i \leq n} (1-y_i)$$
 
 One can thus question the situation where~\mbox{$\lambda^{\star}=1$}. \\

  In an economy with no final expenditure, each pole shares its production with the other ones, the final expenditure state is thus connected to itself with probability~$1$. The transition matrix rank is $2$, with a basis of two eigenvectors for the eigenvalue $1$, corresponding to two recurrent classes. The spectral gap is thus minimized (null). This situation is an utopia.\\

  In the situation of non-null final expenditure, let $\mathbf{v}$ be the non-negative eigenvector with respect to the spectral radius of the substochastic matrix~$\mathbf{A}$, as defined in theorem \ref{Perron-Frobenius}. The vector~$\mathbf{v}$ cannot be constant unless:
  
  $$\forall \, i \,\in\, \left \lbrace 1, \hdots, n \right \rbrace \, : \quad (1-y_i)=(1-y)=\lambda^{\star}$$
  
  \noindent in which case:

\begin{align*}
\sum_{j=1}^n \mathbf{A}_{i j} \mathbf{v}_j&=v \sum_{j=1}^n \mathbf{A}_{i j} =\lambda^{\star} v
\end{align*}

\newpage
  In the situation where~$\mathbf{v}$ is not constant, let us set:
  
   $$i^{\star}=\arg\max_{1 \leq i \leq n} (1-y_i) \quad \text{and} \quad \overline{i}=\arg\max_{1 \leq i \leq n} \mathbf{v}_i$$
   
   One has then:

\begin{enumerate}
\item[\emph{i}.] If $i^{\star}\neq \overline{i}$:

\begin{align*}
\sum_{j=1}^n \mathbf{A}_{\overline{i} j} \mathbf{v}_j&= \lambda^{\star} \sum_{j=1}^n \left(\frac{\mathbf{A}_{\overline{i}j}}{\lambda^{\star}}\right) \mathbf{v}_j <\lambda^{\star} \mathbf{v}_{\overline{i}}
\end{align*}

\noindent since~\mbox{$\displaystyle \sum_{j=1}^n \left(\frac{\mathbf{A}_{\overline{i}j}}{\lambda^{\star}}\right)<1$}, which contradict the spectral identity.

\item[\emph{ii}.] If $i^{\star}= \overline{i}$:

\begin{align*}
\sum_{j=1}^n \mathbf{A}_{\overline{i} j} \mathbf{v}_j&= \lambda^{\star} \sum_{j=1}^n \frac{\mathbf{A}_{\overline{i} j}}{\lambda^{\star}} \mathbf{v}_j =\lambda^{\star} \mathbf{v}_{\overline{i}}
\end{align*}

\noindent which is not possible for a convex combination unless: $$\mathbf{A}_{\overline{i}\overline{i}}=\lambda^{\star} \quad \text{and} \quad \mathbf{A}_{\overline{i}j}=0 \quad \text{for } j\neq \overline{i}$$

\end{enumerate}

\hskip 0.5cm This result describes an \textbf{almost-infinite loop}. This corresponds to an economy where the pole with the least final expenditures rate is in (supply) autarky. The production is then reinjected in the system, and creates spillover effects, while the speed of absorption by the final expenditure is at the slowest level. The pole at stake is the \textbf{transformation} one.\\

  Let us consider now the situation where the whole production is consumed as final expenditure. The transition matrix~$\mathbf{P}$ will have rank $1$, with one eigenvector of the eigenvalue $1$ corresponding to the absorbing class, the other eigenvalues are null and the spectral gap will be maximized (equals to $1$). This situation is also an utopia.\\

  With non-zero final expenditure, we proceed in the same way as before. We set

$$\lambda^{\star}=\displaystyle \min_{1 \leq i \leq n} (1-y_i)$$

\noindent and:

 $$i_{\star}=\displaystyle \arg\min_{1\leq j \leq n} (1-y_i)\quad , \quad \underline{i}=\arg\min_{1\leq i\leq n}  \mathbf{v}_i$$

One has then:

\begin{enumerate}
	\item[\emph{i}.]  If $i_{\star}\neq \underline{i}$:

\begin{align*}
\sum_{j=1}^n \mathbf{A}_{\underline{i} j} \mathbf{v}_j&= \lambda^{\star} \sum_{j=1}^n \left(\frac{\mathbf{A}_{\underline{i}j}}{\lambda^{\star}}\right) \mathbf{v}_j >\lambda^{\star} \mathbf{v}_{\underline{i}}
\end{align*}

\noindent since~\mbox{$\displaystyle \sum_{j=1}^n \left(\frac{\mathbf{A}_{\underline{i}j}}{\lambda^{\star}}\right)>1$}, contradiction.

	\item[\emph{ii}.] If~$i_{\star}= \underline{i}$:

\begin{align*}
\sum_{j=1}^n \mathbf{A}_{\underline{i} j} \mathbf{v}_j&= \lambda^{\star} \sum_{j=1}^n \frac{\mathbf{A}_{\underline{i} j}}{\lambda^{\star}} \mathbf{v}_j =\lambda^{\star} \mathbf{v}_{\underline{i}}
\end{align*}

\noindent which is not possible for a convex combination unless~\mbox{$\mathbf{A}_{\underline{i}\underline{i}}=\lambda^{\star}$} and~\mbox{ $\mathbf{A}_{\underline{i}j}=0$ for $j\neq \underline{i}$}.\\

\end{enumerate}

  To ensure that, for the other eigenvalues~$\lambda_i$, $i=2,\hdots,n$, one has:
  
$$\lambda_i\leq \lambda^{\star}$$

 we apply the maximal spectral radius resonating to the matrix resulting from $\mathbf{A}$ by suppressing the line and the column $\underline{i}$, this matrix is substochastic, so we can apply Perron-Frobenius theorem \ref{Perron-Frobenius} for non-negative matrices to get the condition

\begin{align*}
\min_{i\neq \underline{i}} \sum_{j\neq \underline{i}} \mathbf{A}_{ij}&\leq \lambda^{\star}
\end{align*}

  This solution describes an \textbf{almost-pyramidal structure}, where the pole $i_{\star}$ of the highest final expenditure rate shares the remaining production proportion~\mbox{$\lambda^{\star}$} with himself, one of the other poles at least shares a production proportion not exceeding $\lambda^{\star}$, the surplus $(1-y_i)-\lambda^{\star}$ goes to the pole $i_{\star}$. The spillover effects in this situation is minimized. We will call this pole the \textbf{outlet pole}.\\
\\

  A particular case emerges when the production is fairly shared with the other poles. For~\mbox{$(i,j)\, \in\, \{1,\hdots,n\}^2$}, the~\mbox{$j^{th}$} pole shares a production~\mbox{$\displaystyle\frac{(1-y_i)}{n}$} with the the~\mbox{$i^{th}$} one The matrix $\mathbf{A}$ is singular with rank~$1$, and its spectrum contains only one non-zero eigenvalue, corresponding to:

\begin{align*}
\sum_{j=1}^n \mathbf{A}_{ij} \mathbf{v}_j &= \frac{(1-y_i)}{n} \sum_{j=1}^n \mathbf{v}_j = \lambda^{\star} \mathbf{v}_i \\
\end{align*}

Summation over~$i$ yields:

\begin{align*}
\sum_{i=1}^n\sum_{j=1}^n \mathbf{A}_{ij} \mathbf{v}_j &=\sum_{i=1}^n \frac{(1-y_i)}{n}\,  \sum_{j=1}^n \mathbf{v}_j = \lambda^{\star} \sum_{j=1}^n \mathbf{v}_j \\
\lambda^{\star}&= \sum_{i=1}^n \frac{(1-y_i)}{n}
\end{align*}

  This is a situation of \textbf{fair division}, the connection between poles is maximal, no pole gets a special treatment.

\vskip 1cm

\subsection{Marginal effects and fundamental matrix}

 \hskip 0.5cm Let us consider~\mbox{$(i,j)\, \in\, \left \lbrace 1,\hdots, n \right \rbrace^2$}. We recall the marginal effects identity:

\begin{align*}
\frac{\partial \mathbf{X}_i}{\partial \mathbf{Y}_j}&=\mathbf{O}_{ij}\\
\end{align*}

\noindent or, expressed in terms of an infinite sum

\begin{align*}
\mathbf{O}=\displaystyle \sum_{k=0}^{+\infty}\mathbf{\Theta}^k
\end{align*}

 The minimal marginal effect is:
 $$\mathbf{O}_{ij}=\delta_{ii}(ij) $$
 
 This corresponds to a situation where there is no walk from~$i$ to~$j$. In other words, the output of pole $i$ never integer the production process of pole $j$.\\

  The maximal marginal effect is:
   $$\mathbf{O}_{ii}=\displaystyle \sum_{k=0}^{+\infty}(1-y_i)^k=\displaystyle\frac{1}{y_i} \quad \text{if} \quad  \mathbf{\Theta}_{ij}=(1-y_i) \, \delta_{ii}(ij)
 $$
 
 \noindent  and :
 
 $$\mathbf{O}_{ij}=\displaystyle\sum_{k=0}^{+\infty}(1-y_i)\, (1-y_j)^k=\displaystyle\frac{(1-y_i)}{y_j} \quad \text{for $j\neq i$ if  $\mathbf{\Theta}_{ij}=(1-y_i)$, $\mathbf{\Theta}_{jj}=(1-y_j)$ and $\mathbf{\Theta}_{ik}=\mathbf{\Theta}_{jk}=0$ for $k\neq j$}$$
 
   the output of pole $i$ is totally used in the production of pole $j$, and pole $j$ is in autarky.

\vskip 1cm

\subsection{Production cycle}

\hskip 0.5cm The production process duration is defined by means of the expected time before absorption vector:

\begin{align*}
\mathbf{t}&=\left(
\begin{matrix}
\mathbf{t}_1\\
\vdots\\
\mathbf{t}_{n}
\end{matrix}
\right)=\mathbf{N}\,\mathbf{1}
\end{align*}

\noindent where

\begin{align*}
\mathbf{N}&=(\mathbf{I}-\mathbf{A})^{-1}=\sum_{k=0}^{+\infty}\mathbf{A}^k
\end{align*}

By analogy with previous section, the minimal value:

$$\displaystyle \min_{1 \leq i \leq n}\mathbf{t}_i$$

\noindent  with non-zero final expenditure, is reached when the whole output of the~\mbox{$i^{th}$} pole  is destined to the outlet one in autarky~\mbox{$i_{\star}$}. One has then: $$\underline{\mathbf{t}_i}=\displaystyle 1+\frac{(1-y_i)}{y_{i_{\star}}}\quad \text{if } i\neq i_{\star} \quad \text{or }   \underline{\mathbf{t}_i}=\displaystyle \frac{1}{y_{i_{\star}}} \text{ if }i= i_{\star}$$

  In the same way, the maximal value of $\mathbf{t}_i$, $i=,\hdots,n$, is reached when all the output of the~\mbox{$i^{th}$} pole is destined to the transformation pole $i^{\star}$ in autarky, in which case:

  	$$\overline{\mathbf{t}_i}=\displaystyle 1+ \frac{(1-y_i)}{y_{i^{\star}}} \quad \text{ if }  i\neq i^{\star}$$
  	
  	\noindent or
  	
  	 $$\overline{\mathbf{t}_i}=\displaystyle \frac{1}{y_{i^{\star}}} \quad \text{ if } i= i^{\star}$$

  The proof is obtained by induction, since:~$\mathbf{A}^k=\mathbf{A}^{k-1}\mathbf{A}$, and:
   $$\mathbf{N}=\mathbf{I}+\displaystyle \sum_{k=1}^{+\infty}\mathbf{A}^k$$
   
One has thus:

\begin{align*}
(1-y_i)(1-y_{i_{\star}}) \leq (1-y_i)\, \min_{1\leq l \leq n} \mathbf{A}_{lj} \leq \mathbf{A}^2_{ij}&=\sum_{l=1}^n \mathbf{A}_{il} \, \mathbf{A}_{lj}\\
&=(1-y_i)\, \sum_{l=1}^n \left(\frac{\mathbf{A}_{il}}{(1-y_i)}\right)\mathbf{A}_{lj}\\
&\leq (1-y_i)\, \max_{1\leq l \leq n}  \mathbf{A}_{lj}\leq (1-y_i)(1-y_{i^{\star}})\\
\end{align*}

\noindent and

\begin{align*}
\frac{1-y_i}{y_{i^{\star}}}&\leq \mathbf{N}_{ij}=\sum_{k=0}^{\infty}\mathbf{A}^k_{ij}\leq \frac{1-y_i}{y_{i_{\star}}}\\
\end{align*}

 In order to quantify the relative duration of a production process $i$, we introduce, for~\mbox{$1 \leq i \leq n$}, the ratio

\begin{align*}
0\leq\Delta \mathbf{t}_i&=\frac{\mathbf{t}_i-\underline{\mathbf{t}_i}}{\overline{\mathbf{t}_i}-\underline{\mathbf{t}_i}}\leq 1
\end{align*}

\vskip 1cm

\newpage
\subsection{Discussion}

\hskip 0.5cm The analysis toolbox presented in this section considers the input-output model from two point of view: a first one, through a \textbf{global vision} based on the spectral radius and the relaxation time, measuring the spillover effect and the return time to equilibrium ; a second one, through a \textbf{local vision} defined by the marginal effects and the time to absorption, measuring the interactions between poles and the production processes duration.\\

 The three measures are deeply related: the almost-infinite loop structure describes a maximal spillover effect induced by the maximal spectral radius. Such a situation corresponds to the longest time to absorption, and the maximum self-marginal effect for the transformation pole.\\

 The almost-pyramidal structure is produced when the spillover effect is minimal, with a spectral radius minimal too. In this case, the absorption time reaches its minimum for the outlet pole.\\

  One may summarize those implications as follows:

\begin{align*}
\textbf{Almost-infinite loop} &\Leftrightarrow \forall\, i\,\in\, \left \lbrace 1, \hdots, n \right \rbrace \,  : \quad \mathbf{t}_{i^{\star}} =\overline{\mathbf{t}_i} \Leftrightarrow \max_{i,j} \frac{\partial \mathbf{X}_i}{\partial \mathbf{Y}_j}=\frac{\partial \mathbf{X}_{i^{\star}}}{\partial \mathbf{Y}_{i^{\star}}}=\frac{1}{y_{i^{\star}}} \\ \\
\textbf{Almost-pyramidal structure} &\Rightarrow \forall\, i\, \in\,  \left \lbrace 1, \hdots, n \right \rbrace \, : \quad \mathbf{t}_{i_{\star}} =\underline{\mathbf{t}_i} \Leftrightarrow \frac{\partial \mathbf{X}_{i_{\star}}}{\partial \mathbf{Y}_{i_{\star}}}=\frac{1}{y_{i_{\star}}} \\
\end{align*}

\vskip 1cm

\section{Moroccan input-output table}

\hskip0.5cm Input-output tables describe the inter-industrial flows of goods and services in current prices (USD million), defined according to industry outputs. We hereafter describe next the Moroccan input-output table of 2015, drawn from the OCDE database~\cite{dataOCDE}. It is a matrix flow of 36 poles, which are coded according to (see~Table~\ref{InputOutputTable}):

\vskip 1cm

\begin{table}

\begin{tabular}{|l|M{2cm}|M{2cm}|M{2cm}|M{2cm}|M{2cm}|M{2cm}|}
    \hline
    \textbf{Code} & D01T03 & D05T06 & D07T08 & D09 & D10T12 & D13T15  \tabularnewline
    \hline
    \textbf{Designation} & Agriculture, forestry and fishing & Mining and extraction of energy producing products & Mining and quarrying of non-energy producing products & Mining support service activities & Food products, beverages and tobacco & Textiles, wearing apparel, leather and related products \tabularnewline
    \hline
 \end{tabular}

\vskip 0.5cm

\begin{tabular}{|M{2cm}|M{2cm}|M{2cm}|M{2cm}|M{2cm}|M{2cm}|M{2cm}|}
    \hline
     D16 & D17T18 & D19 & D20T21 & D22 & D23 & D24 \tabularnewline
    \hline
     Wood and products of wood and cork & Paper products and printing & Coke and refined petroleum products & Chemicals and pharmaceutical products & Rubber and plastic products & Other non-metallic mineral products &  Basic metals \tabularnewline
    \hline
 \end{tabular}

\vskip 0.5cm

\begin{tabular}{|M{2cm}|M{2cm}|M{2cm}|M{2cm}|M{2cm}|M{2cm}|M{2cm}|}
    \hline
     D25 & D26 & D27 & D28 & D29 & D30 & D31T33 \tabularnewline
    \hline
     Fabricated metal products & Computer, electronic and optical products & Electrical equipment & Machinery and equipment, nec & Motor vehicles, trailers and semi-trailers & Other transport equipment &  Other manufacturing; repair and installation of machinery and equipment \tabularnewline
    \hline
 \end{tabular}

\vskip 0.5cm

\begin{tabular}{|M{2cm}|M{2cm}|M{2cm}|M{2cm}|M{2cm}|M{2cm}|M{2cm}|}
    \hline
     D35T39 & D41T43 & D45T47 & D49T53 & D55T56 & D58T60 & D61 \tabularnewline
    \hline
     Electricity, gas, water supply, sewerage, waste and remediation services & Construction & Wholesale and retail trade; repair of motor vehicles & Transportation and storage & Accommodation and food services & Publishing, audiovisual and broadcasting activities &  Telecommunications \tabularnewline
    \hline
 \end{tabular}

\vskip 0.5cm

\begin{tabular}{|M{2cm}|M{2cm}|M{2cm}|M{2cm}|M{2cm}|M{2cm}|M{2cm}|}
    \hline
     D62T63 & D64T66 & D68 & D69T82 & D84 & D85 & D86T88 \tabularnewline
    \hline
     IT and other information services & Financial and insurance activities & Real estate activities &  Other business pole services & Public admin. and defense; compulsory social security & Education &  Human health and social work \tabularnewline
    \hline
 \end{tabular}

\vskip 0.5cm

\begin{tabular}{|M{2cm}|M{2cm}|}
    \hline
     D90T96 & D97T98  \tabularnewline
    \hline
     Arts, entertainment, recreation and other service activities & Private households with employed persons  \tabularnewline
    \hline
 \end{tabular}
 
 \caption{The Moroccan input-output table of 2015.}
 \label{InputOutputTable}
 \end{table}
\vskip 1cm

\hskip 0.5cm The input-output table reflects the supply-uses identity:\\

\begin{align*}
\overset{\textbf{Supply}}{\overbrace{\textbf{Output} + \textbf{Imports}}} &= \overset{\textbf{Uses}}{\overbrace{\textbf{Inter. consump.} + \underset{\textbf{Final expenditure}}{\underbrace{\textbf{Domestic consump.} + \textbf{GFCF} + \Delta \textbf{ inventories} + \textbf{Exports}}}}}
\end{align*}

Let us recall that the indirect orientation leads to a Markov chain with transition matrix~\mbox{$\widehat{\mathbf{A}}$ }, which represents the repartition of the supply flows. The direct (use/demand) orientation for which the monetary dual leads to a Markov chain with transition matrix $\widehat{\mathbf{G}}=\widehat{\mathbf{\Theta}}^T$, which represents the repartition of the uses flows.\\

\begin{figure}[h!]
		\begin{center}
		\includegraphics[scale=1]{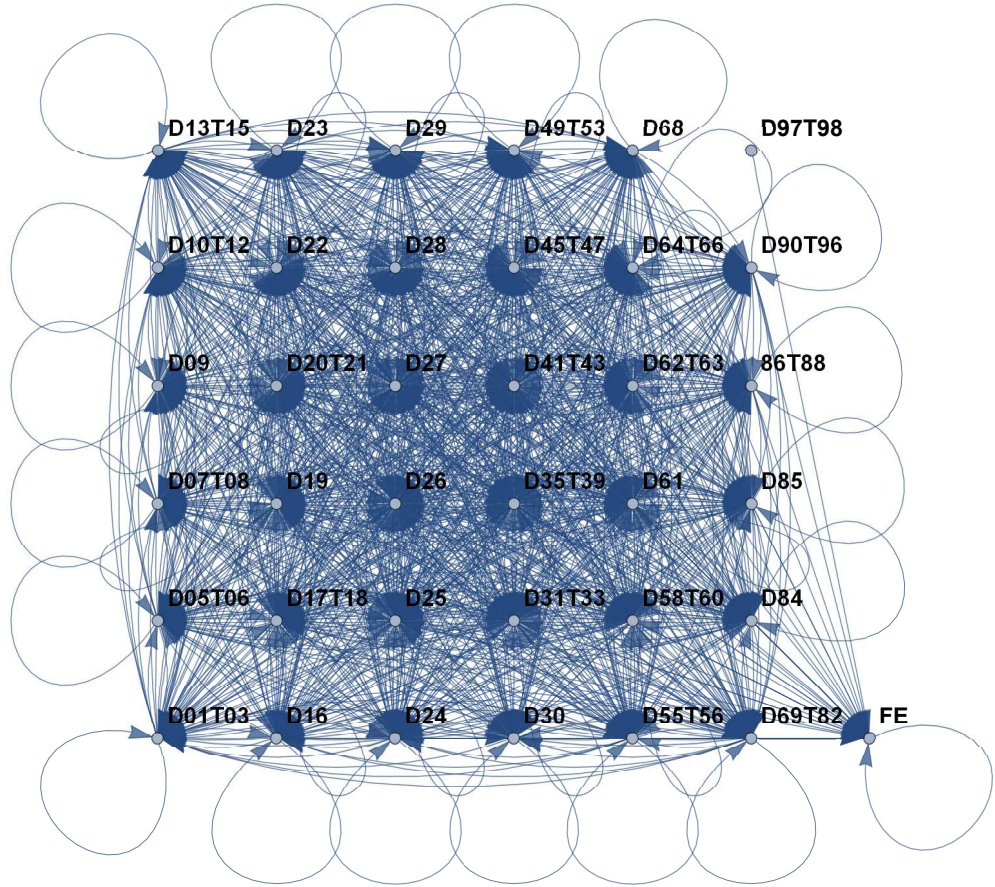}
			\captionof{figure}{Flows web.}
			\label{Web}
		\end{center}
	\end{figure}

 We hereafter plot the Markov chain's web~(see Figure~\ref{Web}).

	\vskip 1cm
	
	\vskip 1cm

  The web contains three strong components: the final expenditure (F.E.) representing the absorbent state, D97T98 for private households with employed persons, and the other poles. All the poles are transient states communicating and sharing production, except for the  private households pole with employed persons, totally destined to final expenditure.\\

\begin{figure}[!htb]
		\begin{center}
		\includegraphics[scale=1]{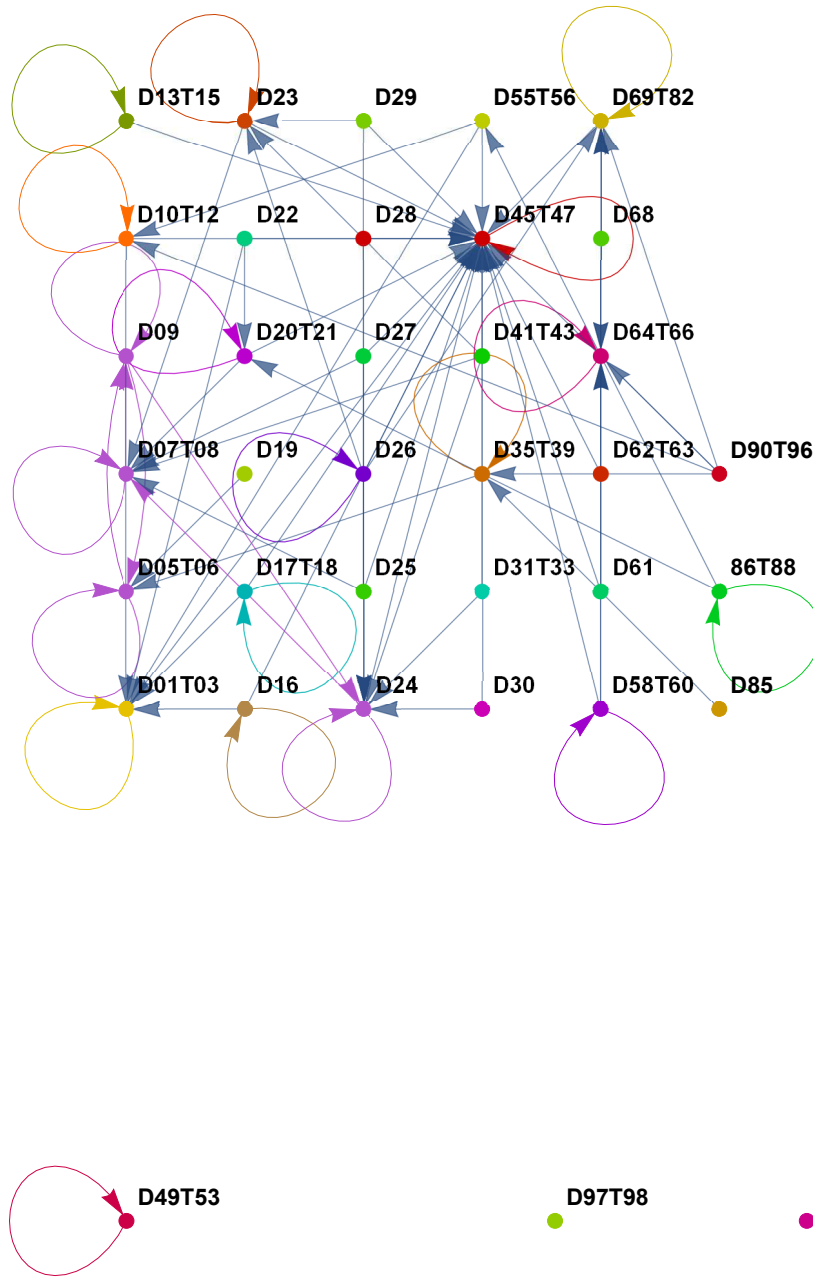}
			\captionof{figure}{Essential flows of the direct orientation.}
			\label{DemandWeb}
		\end{center}
	\end{figure}

For a better understanding of connections between poles, we have choosen to keep only connections that exceed~$\displaystyle\frac{1}{37}$ (the fair division case). This yield  the graphs~\ref{DemandWeb}:

	\vskip 1cm
	
	\vskip 1cm
	
	\begin{figure}[!htb]
		\begin{center}
		\includegraphics[scale=1]{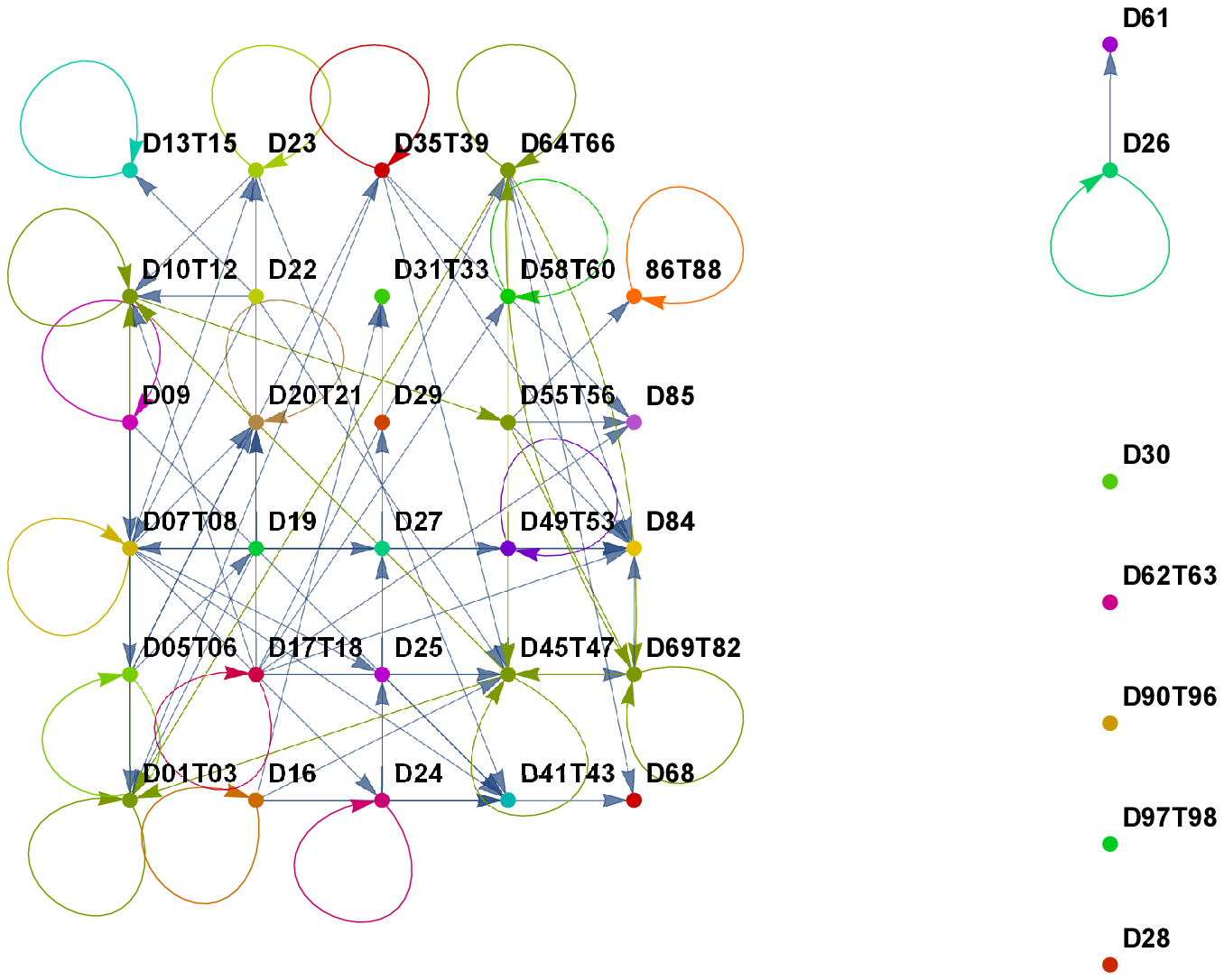}
			\captionof{figure}{Essential flows of the indirect orientation.}
			\label{SupplyWeb}
		\end{center}
	\end{figure}
	
	\vskip 1cm

 The essential strong components of the indirect (supply) orientation are all singletons, except for the supply group $\{$D01T03 : Agriculture, forestry and fishing, D10T12 : Food products, beverages and tobacco, D45T47 : Wholesale and retail trade; repair of motor vehicles, D55T56 : Accommodation and food services, D64T66 : Financial and insurance activities, D69T82 : Other business pole services$\}$ representing the food industry.\\

 \hskip 0.5cm  From the user's point of view, we found a different essential component represented by the uses group $\{$D05T06 : Mining and extraction of energy producing products, D07T08 : Mining and quarrying of non-energy producing products, D09 : Mining support service activities, D24 : Basic metals$\}$ representing the metallurgy industry.\\

Table~\ref{SpectralProperties} summarizes the spectral properties of the technical (and trade) matrix: spectral radius, the relaxation time, the minimal and maximal final expenditure rates :

\vskip 1cm

\begin{center}
	\begin{table}
\begin{tabular}[!htb]{|c|c|c|c|}
    \hline
    $\lambda^{\star}$ &  $0.293934$ \tabularnewline
    \hline
    $t_{\text{rel}}$ & $1.4163$ \tabularnewline
    \hline
    $1-y_{\underline{i}}$ & $0.911226$ \tabularnewline
    \hline
    $1-\overline{y}$ & $0.308206$ \tabularnewline
    \hline
    $1-y_{\overline{i}}$ & $0$ \tabularnewline
    \hline
 
\end{tabular}

\caption{The spectral properties of the technical (and trade) matrix.}
    \label{SpectralProperties}
    
    \end{table}

\end{center}

\vskip 1cm

 Those parameters provide a global appreciation of the spillover effects, and of the attenuation speed. In order to give a relative evaluation of them, in terms of the economy potential, we introduce a new measure,~$f$, taking the values:

$$ 
f(1-y_{i^{\star}})= 1 \quad , \quad 
f(1-\overline{y}) =\displaystyle  \frac{1}{2}  \quad , \quad 
f(1-y_{i_{\star}}) = 0\\
$$

 The value~$1$ occurs in the situation of almost-infinite loop with maximum spillover effect, $0$ occurs in the situation of almost-pyramidal structure with the minimum spillover effect, while~\mbox{$0.5$} is the intermediary situation of fair division. $f$ is defined using Lagrange interpolation as:

\begin{align*}
f(x)&=\frac{1}{2}\frac{(x-a)(x-b)}{(c-a)(c-b)}+\frac{(x-a)(x-c)}{(b-a)(b-c)}\\
\end{align*}

\noindent where:

$$a=1-y_{i^{\star}} \quad , \quad b=1-y_{i_{\star}} \quad , \quad c=1-\overline{y}$$

  In the Moroccan case, one has:
  $$f(\lambda^{\star})=0.480498$$

  \noindent which corresponds to a situation of downward fair division, characterized by weak pyramidal structure. The spillover effect of the economy is medium. One may note that the relaxation time indicated~\mbox{$1.4163$} steps before the system could return to the steady state of no production.\\

  From a local point of view, the expected time before absorption (E.T.A.) represents the duration of an industrial production process before been absorbed by final expenditure. We summarize in Table~\ref{ExpectedTimeBeforeAbsorption} the expected time before absorption~\mbox{$\mathbf{t}_i$} when starting in state~$i$.

\vskip 1cm

\begin{table}[!htb]
\begin{tabular}{|c|c|c|c|c|c|}
    \hline
     \textbf{Pole} & \textbf{E.T.A.} & $\overline{\mathbf{t}_i}$ & $\underline{\mathbf{t}_i}$ & $\Delta{\mathbf{t}_i}$ \tabularnewline
    \hline
     \textbf{D01T03} & $1.59156$ & $5.798013421$ & $1.42593988$ & $0.037881366$ \tabularnewline
    \hline
     \textbf{D05T06} & $2.60198$ & $11.26453202$ & $1.911225784$ & $0.073851342$ \tabularnewline
    \hline
     \textbf{D07T08} & $1.95377$ & $8.258599566$ & $1.644376487$ & $0.046777$ \tabularnewline
    \hline
     \textbf{D09} & $2.83355$ & $10.953322577$ & $1.88359841$ & $0.104738752$ \tabularnewline
    \hline
     \textbf{D10T12} & $1.20384$ & $2.703975775$ & $1.151269114$ & $0.033857578$ \tabularnewline
    \hline
     \textbf{D13T15}  & $1.18072$ & $2.66411798$ & $1.147730769$ & $0.02175515$ \tabularnewline
    \hline
     \textbf{D16}  & $1.74648$ & $7.84390081$ & $1.60756193$ & $0.022275581$ \tabularnewline
    \hline
     \textbf{D17T18} & $2.01144$ & $8.951063463$ & $1.705849426$ & $0.042178267$ \tabularnewline
    \hline
     \textbf{D19} &  $1.75013$ & $6.820058155$ & $1.5166711$ & $0.044020717$ \tabularnewline
    \hline
     \textbf{D20T21} &  $1.18653$ & $2.571492976$ & $1.139508057$ & $0.032836898$ \tabularnewline
    \hline
     \textbf{D22} & $1.55758$ & $5.52228095$ & $1.401461946$ & $0.0378852$ \tabularnewline
    \hline
     \textbf{D23} & $2.04427$ & $10.824915992$ & $1.872199216$ & $0.019219952$ \tabularnewline
    \hline
     \textbf{D24} & $1.98661$ & $9.384341624$ & $1.744313355$ & $0.031714103$ \tabularnewline
    \hline
     \textbf{D25} & $1.20291$ & $2.745771947$ & $1.154979536$ & $0.03012993$ \tabularnewline
    \hline
     \textbf{D26} & $1.23853$ & $3.136547068$ & $1.189670291$ & $0.025096457$ \tabularnewline
    \hline
     \textbf{D27} & $1.06766$ & $1.638851502$ & $1.056713541$ & $0.018803891$ \tabularnewline
    \hline
     \textbf{D28} & $1.08302$ & $1.646084891$ & $1.05735568$ & $0.043592741$ \tabularnewline
    \hline
     \textbf{D29} & $1.04159$ & $1.353595176$ & $1.031390135$ & $0.031656443$ \tabularnewline
    \hline
     \textbf{D30} & $1.05987$ & $1.56508697$ & $1.050165153$ & $0.018847225$ \tabularnewline
    \hline
     \textbf{D31T33} & $1.17029$ & $2.436422272$ & $1.127517261$ & $0.03267826$ \tabularnewline
    \hline
     \textbf{D35T39} & $1.7901$ & $7.09122291$ & $1.540743539$ & $0.044925212$ \tabularnewline
    \hline
     \textbf{D41T43} & $1.01321$ & $1.107244285$ & $1.009520527$ & $0.037754102$ \tabularnewline
    \hline
     \textbf{D45T47} & $1.58799$ & $5.739340748$ & $1.42073126$ & $0.038729767$ \tabularnewline
    \hline
     \textbf{D49T53} & $1.38415$ & $4.00360591$ & $1.26664276$ & $0.042933439$ \tabularnewline
    \hline
     \textbf{D55T56} & $1.38708$ & $4.263278233$ & $1.289694967$ & $0.032750061$ \tabularnewline
    \hline
     \textbf{D58T60} & $1.15296$ & $2.359375424$ & $1.120677488$ & $0.02606165$ \tabularnewline
    \hline
     \textbf{D61} & $1.07444$ & $1.660539484$ & $1.058638875$ & $0.02625205$ \tabularnewline
    \hline
     \textbf{D62T63} & $1.09222$ & $1.78142887$ & $1.069370735$ & $0.032089044$ \tabularnewline
    \hline
     \textbf{D64T66} & $1.94122$ & $7.789725066$ & $1.60275252$ & $0.054706478$ \tabularnewline
    \hline
     \textbf{D68} & $1.17236$ & $2.43074642$ & $1.127013392$ & $0.034782127$ \tabularnewline
    \hline
     \textbf{D69T82} & $1.52432$ & $5.228638776$ & $1.375394093$ & $0.038649481$ \tabularnewline
    \hline
     \textbf{D84} & $1.03745$ & $1.31337571$ & $1.027819683$ & $0.03372479$ \tabularnewline
    \hline
     \textbf{D85} & $1.03546$ & $1.315536489$ & $1.028011504$ & $0.025905559$ \tabularnewline
    \hline
     \textbf{86T88} & $1.18021$ & $2.531030273$ & $1.135916012$ & $0.031749362$ \tabularnewline
    \hline
     \textbf{D90T96} &  $1.27396$ & $3.286460329$ & $1.202978723$ & $0.034068588$ \tabularnewline
    \hline
     \textbf{D97T98} & $1$ & $1$ & $1$ & $\_$ \tabularnewline
    \hline
    
    \end{tabular}

 \caption{ }
\label{ExpectedTimeBeforeAbsorption}

\end{table}

\vskip 1cm

 The minimal expected time before absorption correspond to the pole D97T98: private households with employed persons whose production is totally oriented to final consumption, the maximal expected time before absorption is about $2.83355$ and correspond to D09: Mining support service activities, the duration of the industrial transformation process for this pole is the longest, and the relative duration ratio is the highest for this pole too.\\

The highest vs lowest values of the sensitivity parameters are summarized in Table~\ref{SensitivityParameters}:

\vskip 1cm

\begin{table}

\begin{tabular}[!htb]{|c|c|c|c|}
    \hline
   $\max_{i,j} \displaystyle\frac{\partial \mathbf{X}_i}{\partial \mathbf{Y}_j}$   &  \textbf{Value} & \textbf{Origin $j$} & \textbf{Target $i$} \tabularnewline
    \hline
  $\mathbf{1}$  &  $1.27355$ & \textbf{D64T66} & \textbf{D64T66} \tabularnewline
    \hline
     $\mathbf{2}$ & $1.21066$ & \textbf{D01T03} & \textbf{D01T03} \tabularnewline
     \hline
     $\mathbf{3}$ & $1.12336$ & \textbf{D09} & \textbf{D09} \tabularnewline
     \hline
     $\mathbf{4}$ & $1.11943$ & \textbf{D23} & \textbf{D23} \tabularnewline
     \hline
     $\mathbf{5}$ & $1.1147$ & \textbf{D35T39} & \textbf{D35T39} \tabularnewline
     \hline
 
\end{tabular}

\vskip 0.5cm

\begin{tabular}[!htb]{|c|c|c|c|}
    \hline
   $\min_{i,j} \displaystyle\frac{\partial \mathbf{X}_i}{\partial \mathbf{Y}_j}$   &  \textbf{Value} & \textbf{Origin $j$} & \textbf{Target $i$} \tabularnewline
    \hline
     $\mathbf{1}$ & $0$ & \textbf{D97T98} & \textbf{Any destination except D97T98} \tabularnewline
     &  & \textbf{Any origin except D97T98} & \textbf{D97T98} \tabularnewline
    \hline
     $\mathbf{2}$ & $9.1386*10^{-6}$ & \textbf{D68} & \textbf{D30} \tabularnewline
     \hline
     $\mathbf{3}$ & $0.0000142119$ & \textbf{D09} & \textbf{D30} \tabularnewline
     \hline
     $\mathbf{4}$ & $0.0000152982$ & \textbf{D28} & \textbf{D30} \tabularnewline
     \hline
     $\mathbf{5}$ & $0.000016291$ & \textbf{D85} & \textbf{D30} \tabularnewline
     \hline
      
\end{tabular}

\caption{The highest vs lowest values of the sensitivity parameters.}
    \label{SensitivityParameters}
\end{table}
\vskip 1cm

 The maximal marginal impacts are obtained for self-sensitivity, we sort them  in decreasing order: D64T66: Financial and insurance activities, D01T03: Agriculture, forestry and fishing, D09 : Mining support service activities, D23: Other non-metallic mineral products, D35T39: Electricity, gas, water supply, sewerage, waste and remediation services.\\

  The minimal marginal impacts is null between D97T98: Private households with employed persons, and the others poles. Then we found the final expenditure marginal impact on D3 : Other transport equipment, from the production of D68: Real estate activities, D09: Mining support service activities, D28 : Machinery and equipment, nec and D85: Education.

\vskip 1cm

\subsection{Benchmark}

 \hskip 0.5cm To enrich our study, we enclose a benchmark panel between: Brazil (BRA), China (CHI), France (FRA), Germany (GER),  Morocco (MOR), Saoudian Arabia (SAO), South Africa (SAF), Thailand (THA), Tunisia (TUN), Turkey (TUR), USA, Vietnam (VIE), increasingly ranged by growth rate~$r$:

\vskip 1cm

\begin{tabular}{|c|c|c|c|c|c|c|}
    \hline
    \textbf{Country} &  \textbf{BRA} &  \textbf{FRA} & \textbf{TUN} & \textbf{SAF} &  \textbf{GER} & \textbf{USA} \tabularnewline
    \hline
    $r$ &  $-3,546$ & $1.113$ & $1.166$ & $1,194$ &  $1.492$ & $2.706$ \tabularnewline
    \hline
    $\lambda^{\star}$ &  $0.4631$ & $0.4186$ & $0.3575$ & $0.4693$ &  $0.3946$ & $0.3860$ \tabularnewline
    \hline
    $t_{\text{rel}}$ & $1.8623$ & $1.7199$ & $1.5564$ & $1.8843$ &  $1.6517$ & $1.6286$ \tabularnewline
    \hline
    $1-y_{\underline{i}}$ & $0.8820$ & $0.8411$ & $0.9252$ & $0.9970$ &  $0.7952$ & $0.8848$ \tabularnewline
    \hline
    $1-\overline{y}$ & $0.4700$ & $0.3441$ & $0.3550$ & $0.4716$ &  $0.3797$ & $0.4552$ \tabularnewline
    \hline
    $1-y_{\overline{i}}$ & $0$ & $0$ & $0$ & $0$ & $0$ & $0$ \tabularnewline
    \hline
    $f(\lambda^{\star})$ & $0.492117$ & $0.591679$ & $0.503008$ & $0.497679$ & $0.518782$ & $0.422013$ \tabularnewline
    \hline
    $\max t_i$ & $2.9190$ & $2.3570$ & $2.9677$ & $3.2989$ & $2.4036$ & $3.1479$ \tabularnewline
    \hline
    $\min t_i$ & $1$ & $1$ & $1$ & $1$ & $1$ & $1$ \tabularnewline
    \hline
    $\arg \max t_i$ & D09 & D64T66 & D09 & D09 & D07T08 & D09 \tabularnewline
    \hline
    $\arg \min t_i$ & D97T98 & D97T98 & D97T98 & D97T98 & D97T98 & D97T98  \tabularnewline
    \hline
\end{tabular}

\vskip 0.5cm

\begin{tabular}{|c|c|c|c|c|c|c|}
    \hline
    \textbf{Country} & \textbf{THA} & \textbf{SAO} & \textbf{MOR} & \textbf{TUR} &  \textbf{VIE} &  \textbf{CHI} \tabularnewline
    \hline
    $r$ &  $3.134$ &  $4.106$ &  $4.536$ & $6.085$ &  $6.679$ &  $7.041$\tabularnewline
    \hline
    $\lambda^{\star}$ & $0.3969$ &  $0.2636$ & $0.2939$ & $0.4780$ &  $0.469$ &  $0.5525$ \tabularnewline
    \hline
    $t_{\text{rel}}$ & $1.6580$ &  $1.3580$ & $1.4163$ & $1.9159$ &  $1.8843$ &  $2.2344$\tabularnewline
    \hline
    $1-y_{\underline{i}}$ & $0.9967$ & $0.9250$ & $0.9112$ & $0.9969$ &  $0.9970$ &  $1$ \tabularnewline
    \hline
    $1-\overline{y}$ & $0.3885$ & $0.2928$ & $0.3082$ & $0.4616$  &  $0.4716$ &  $0.5937$ \tabularnewline
    \hline
    $1-y_{\overline{i}}$ & $0$ &  $0$ & $0$ & $0$  &  $0$ &  $0.0165$ \tabularnewline
    \hline
    $f(\lambda^{\star})$ & $0.509256$ & $0.457765$ & $0.480459$ & $0.516592$ & $0.497376$ & $0.456129$ \tabularnewline
    \hline
    $\max t_i$ & $3.2940$ & $2.3280$ & $2.83355$ & $3.1539$  &  $3.2989$ &  $4.2133$\tabularnewline
    \hline
    $\min t_i$ & $1$ & $1$ & $1$ & $1$  &  $1$ &  $1$ \tabularnewline
    \hline
    $\arg \max t_i$ & D05T06 & D62T63  & D09 & D35T39  &  D09 &  D05T06 \tabularnewline
    \hline
    $\arg \min t_i$ & D97T98 & D97T98 & D97T98 & D97T98  &  D97T98 & D97T98 \tabularnewline
    \hline
\end{tabular}

\vskip 1cm

 The dominance/long run effects measure $f$ brings out three categories in the panel:

\begin{enumerate}
\item[\emph{i}.] \textbf{A weak pyramidal structure:} the measure for those countries are slightly smaller  than~\mbox{$0.5$}, which concern USA, Saoudian Arabia, Morocco, China.
\item[\emph{ii}.]\textbf{Fair division countries:} Brazil, Tunisia, South Africa, Thailand, Vietnam.
\item[\emph{iii}.] \textbf{A weak loop structure:} France, Germany, Turkey.
\end{enumerate}

\begin{figure}[!htb]
		\begin{center}
		\includegraphics[scale=0.75]{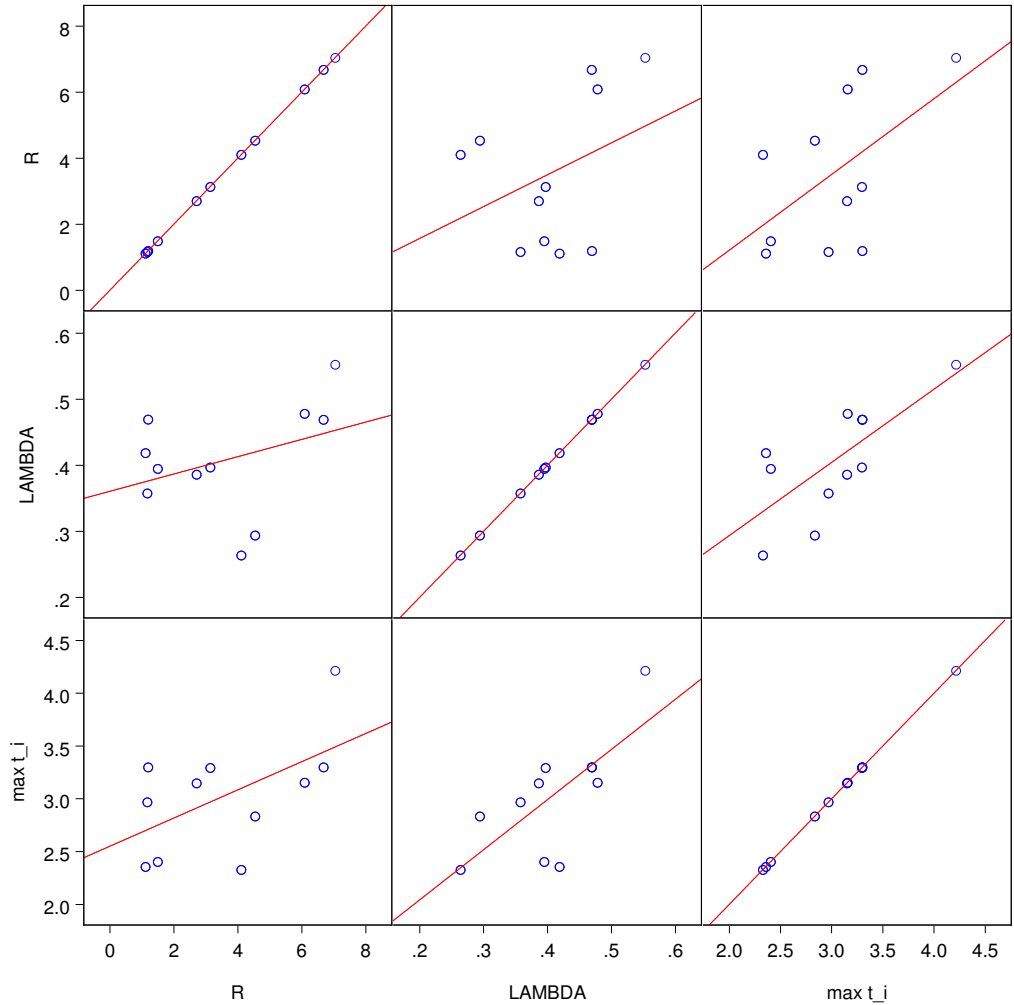}
			\captionof{figure}{Scatterplots.}
			\label{Scatterplot}
		\end{center}
	\end{figure}

 We now propose to question the possible relation between the growth rate~$r$, the spectral radius~\mbox{$\lambda^{\star}$}, and the longest time of absorption~\mbox{$\max t_i$} for these panel of countries, as displayed in Figure\ref{Scatterplot}:

	\vskip 1cm
	
	\vskip 1cm

The correlation and the corresponding p-value of the t-test are reported in Table~\ref{Correlation}, wherecomputations have been done after putting apart the outlier corresponding to Brazil:

\vskip 1cm

\begin{tabular}[!htb]{|c|c|c|c|}
    \hline
   \textbf{Correlation}   &  $r$ & $\lambda^{\star}$ & $\max t_i$ \tabularnewline
    \hline
  $r$  &  $1$ & $0.355589 (0.2832)$ & $0.553374 (0.0774)$ \tabularnewline
    \hline
     $\lambda^{\star}$ & $0.355589 (0.2832)$ & $1$ & $0.725540 (0.0115)$ \tabularnewline
     \hline
     $\max t_i$ & $0.553374 (0.0774)$ & $0.725540 (0.0115)$ & $1$ \tabularnewline
     \hline
\end{tabular}
     \label{Correlation}
\vskip 1cm

This shows a significant positive relation between the maximal production process duration, and the growth rate, the relation is positive too with the spectral radius, but it not seem to be statistically significant.

\vskip 1cm

\section{Conclusion}

\hskip 0.5cm This paper had a threefold purpose. First, we aimed at introducing a local measure of dominance based on the spectral properties of the input-output matrix. This has highlighted a local-global duality, in so far as the determinant is nothing more than the product of the eigenvalues.\\

Second, we wanted to reconcile sensitivity analysis and dominance theory - which was not possible without a local measure of dominance.\\

Third, we wished to give a new lecture of the input-output matrix in term of Markov chains, with a deeper interpretation of the underlying dynamics. One must of course bear in mind that the classical lecture is reduced to the analysis of mechanical transition, and does not enable one to determine the intrinsic and fundamental properties of the production process: interdependency and long run effects.\\

With respect to Lantner's classification evoked at the beginning of our study, we thus place ourselves in the line of the first category of contributions, where we plan to bring further developments.

\bibliographystyle{alpha}
\bibliography{BibliographieLebert}

\end{document}